\documentclass[12pt]{article}
\usepackage{amsthm,amsmath, graphicx, booktabs, multirow, xcolor, amssymb, tikz, adjustbox, epstopdf, upgreek, algorithmic}
\usepackage{latexsym}
\usepackage{graphicx}
\usepackage{enumitem}
\usepackage{fullpage}
\usepackage{setspace}
\usepackage{longtable}
\RequirePackage{natbib}
\bibpunct{(}{)}{;}{a}{}{,}
\usepackage{dcolumn}
\newcolumntype{.}{D{.}{.}{-1}}
\newcolumntype{d}[1]{D{.}{.}{#1}}
\usepackage[top=.8in, left=.5in, right=.5in,
bottom=.8in]{geometry}
\usepackage[compact]{titlesec}
\usepackage{booktabs}
\usepackage{url} 

\usepackage{color}
\RequirePackage[colorlinks,citecolor=blue,urlcolor=blue]{hyperref}

\newtheorem{remark}{Remark}
\newtheorem{lemma}{Lemma}
\newtheorem{theorem}{Theorem}
\newtheorem{assumption}{Assumption}

\newcommand{\bm}{\boldsymbol}

\newcommand{\bX}{\boldsymbol{X}}

\usepackage{caption}
\usepackage{subcaption}
\usepackage[linesnumbered, ruled, vlined]{algorithm2e}

\expandafter\def\expandafter\normalsize\expandafter{\normalsize\setlength\abovedisplayskip{0pt}}
\expandafter\def\expandafter\normalsize\expandafter{\normalsize\setlength\belowdisplayskip{0pt}}
\expandafter\def\expandafter\normalsize\expandafter{\normalsize\setlength\abovedisplayshortskip{0pt}}
\expandafter\def\expandafter\normalsize\expandafter{\normalsize\setlength\abovedisplayshortskip{0pt}}

\begin{document}
\pagestyle{plain}

\def\spacingset#1{\renewcommand{\baselinestretch}%
{#1}\small\normalsize} \spacingset{1.5}


  \title{Robust and flexible learning of a high-dimensional classification rule using auxiliary outcomes}
  \author{Muxuan Liang \thanks{
  	Department of Biostatistics, University of Florida, Gainesville, Florida, U.S.A. }\\
  	\and
  	Jaeyoung Park \thanks{
  	Booth School of Business, University of Chicago, Chicago, Illinois, U.S.A. }\\
	\and
	Qing Lu \thanks{
  	Department of Biostatistics, University of Florida, Gainesville, Florida, U.S.A. }\\
  	\and
  	Xiang Zhong\thanks{
  	Department of Industrial and Systems Engineering, University of Florida, Gainesville, Florida, U.S.A. }}
\date{}  	
  	
  \maketitle

\thispagestyle{empty}
\abstract{
Correlated outcomes are common in many practical problems. In some settings, one outcome is of particular interest, and others are auxiliary. To leverage information shared by all the outcomes, traditional multi-task learning (MTL) minimizes an averaged loss function over all the outcomes, which may lead to biased estimation for the target outcome, especially when the MTL model is mis-specified. In this work, based on a decomposition of estimation bias into two types, within-subspace and against-subspace, we develop a robust transfer learning approach to estimating a high-dimensional linear decision rule for the outcome of interest with the presence of auxiliary outcomes. The proposed method includes an MTL step using all outcomes to gain efficiency, and a subsequent calibration step using only the outcome of interest to correct both types of biases. We show that the final estimator can achieve a lower estimation error than the one using only the single outcome of interest. Simulations and real data analysis are conducted to justify the superiority of the proposed method.
}

\newcommand{\n}{\noindent}

\newpage
\section{Introduction}
\label{sec:intro}

With the adoption of electronic health record systems, datasets increasingly massive in volume and diverse in variable category have been leveraged for knowledge discovery and clinical decision support. In some datasets, in addition to the patient outcome of primary interest, multiple relevant health outcomes are available. In this paper, we denote these relevant outcomes as auxiliary outcomes, and our goal is to study how to safely use these auxiliary outcomes to help predict a binary target outcome in a classification framework with high-dimensional linear decision rules.

Our motivating example is predicting whether the improvement in hip joint functions fails to achieve the minimal clinical importance difference (MCID) after total hip arthroplasty (THA). The Hip disability and Osteoarthritis Outcome Scores for Joint Replacement (HOOS JR) survey is a common instrument to measure THA patients' health outcomes. Predicting whether the change of the overall score  measured in preoperative and postoperative surveys exceeds the MCID can help inform whether surgery is necessary. Besides the overall score, there are also other related questionnaire items such as the disease-specific questionnaires that measure the improvement in various perspectives, including pain, sleep, fatigue, and function \citep{kunze2020development, katakam2022development}. This motivates us to explore, whether we can leverage these related questionnaire items -- auxiliary outcomes to facilitate target outcome prediction (i.e., whether the overall score change exceeds MCID).

To model related outcomes jointly, multi-task learning (MTL) has emerged, aiming to exploit commonalities and differences across outcomes \citep{caruana1997multitask}. In MTL, it is  typically assumed that some parameters are similar across tasks \citep{Bakker2004, Yu2005, Ando2005, Zhang2008, argyriou2007spectral, maurer2013sparse, zhu2011infinite, titsias2011spike}, or these tasks bear a shared sparsity structure \citep{obozinski2008high, Lounici2009, Yang2009, rao2013sparse, gong2014efficient, gong2013multi, wang2016multiplicative}. Subsequently, a common feature representation can be learned through MTL, and this approach has been widely applied in many fields \citep{zhang2014facial, liu2015multi, zhang2016deep, mrkvsic2015multi, li2014heterogeneous, shinohara2016adversarial, liu2017adversarial}. In MTL, since outcomes are equally important, the objective function to be minimized is the averaged loss overall tasks. 
Different from MTL, we only address the performance of predicting the target outcome. The decision rule learned in MTL driven by the averaged loss might be biased towards predicting the auxiliary outcomes rather than the target outcome, i.e., the jointly learned decision rule may not perform well when predicting the target outcome. Thus, our objective is to develop a robust learning approach that is capable of exploiting commonalities and differences across outcomes with guaranteed performance in target outcome prediction.

Focusing on the performance of predicting the target outcome, a commonly used approach is transfer learning \citep{olivas2009handbook}. Transfer learning aims to improve the performance of target learners on target domains by transferring the knowledge contained in different but related source domains \citep{zhuang2020comprehensive}. Recently, \cite{li2020transfer} and \cite{bastani2021predicting} addressed transfer learning problems in high-dimensional linear regression; \cite{tian2022transfer} addressed transfer learning problems in high-dimensional generalized linear models. In their proposed procedures, they 1) adopt a common working model for all auxiliary outcomes; 2) assume the contrast between the parameters in the target model and those in the auxiliary models are sufficiently close in $l_1$ or $l_0$ norm. 

However, these assumptions are easily violated in many practical settings such as our motivating example. First, the auxiliary outcomes are related but different, and thus, they are not likely to share exactly the same model. Second, the requirement regarding the contrast between the parameters in the target model and those in the auxiliary models can be restrictive for classification problems. For instance, considering both the target and auxiliary outcomes follow logistic regression models, if the parameters in the target model are twice as large as those in the auxiliary models, the contrast of the two sets of parameters is not necessarily small in $l_1$ or $l_0$ norm. However, from the perspective of classification problems, the optimal decision boundaries are identical for the target and auxiliary outcomes. Thus, there is a need for a more flexible learning approach that efficiently utilizes the possible similarity between decision boundaries, rather than focusing on the contrast of parameters, especially for classification problems.

In this work, we develop a robust and flexible learning approach using auxiliary outcomes to aid the estimation of a high-dimensional linear decision rule for the target outcome. Specifically, we propose a two-stage procedure. In the first stage, a common linear representation of the covariates is learned with all auxiliary outcomes using MTL to gain efficiency by borrowing relevant information from auxiliary outcomes. In the second stage, a calibration procedure is performed to reduce or correct the bias induced in the first stage to ensure the robustness of the estimator for the target outcome prediction. The candidate estimated decision rules are constructed from this calibration procedure. Furthermore, we use a cross-fitting procedure to consistently select the estimate with the lowest estimation error among all candidate estimated decision rules. Compared with the existing literature, our contributions are the following. In the first stage, different from \cite{li2020transfer, tian2022transfer}, where the working models for auxiliary outcomes share similar coefficients and intercepts, we posit different decision rules (or models) for different outcomes to accommodate possible heterogeneity. 
In the second stage, instead of assuming that the contrast between the parameters in models for auxiliary outcomes and the target outcome enjoys a small $l_1$ norm or a sparse $l_0$ norm, we define a novel concept of within-subspace bias and against-subspace bias, and we only assume that the minimal against-subspace bias is sparse in $l_0$ norm or small in $l_1$ norm, which is a weaker condition than those in \cite{li2020transfer, bastani2021predicting, tian2022transfer}. Theoretically, we show that the proposed estimator can achieve a lower estimation error than that using only the target outcome, even if the conditions in \cite{li2020transfer, tian2022transfer} are violated. Especially, we show that with the presence of many weakly dependent outcomes, our proposed method can also lead to a convergence rate faster than the derived rate in \cite{li2020transfer,bastani2021predicting,tian2022transfer} and faster than using only the target outcome.

The rest of the paper is organized as follows. Section 2 introduces the proposed method. In Section 3, we investigate the theoretical properties of the proposed method. In Section 4, we conduct simulations to compare our method with other methods, especially MTL and methods in \cite{li2020transfer}. In Section 5, we apply the proposed method to the motivating study for THA patients. We present a discussion and concluding remarks in Section 6.

\section{Robust and flexible learning using auxiliary outcomes with possible heterogeneous models}
\label{sec:method}

Let $\bX\in \mathbb{R}^p$ be a $p$-dimensional covariate and $Y_0\in \left\{1, -1\right\}$ be a univariate target outcome. We assume that some auxiliary outcomes are available along with the target outcome $Y_0$. We denote the auxiliary outcomes as $Y_1$, $Y_2$, $\cdots$, $Y_J$ $\in \left\{1, -1\right\}$, where $J$ is the number of auxiliary outcomes.

In our motivating example, the target outcome and auxiliary outcomes are available in the same dataset. There are other scenarios where the target outcome and auxiliary outcomes are not in the same dataset. For example, we may have a separate dataset containing only the auxiliary outcomes and covariates, denoted as the source-only dataset. To accommodate this scenario, we assume that we observe $n$ samples in the target dataset where both the target outcome and auxiliary outcomes are available, i.e., $\left\{(\bX_i, Y_{0,i}, Y_{1,i}, \cdots, Y_{J,i})\right\}_{i=1}^n$; in addition, we also observe $N-n$ samples in the source-only dataset where only auxiliary outcomes are available, i.e., $\left\{(\bX_i, Y_{1,i}, \cdots, Y_{J,i})\right\}_{i=n+1}^N$. We use $R_i=0$ to indicate samples coming from the target dataset, and $R_i=1$, from the source-only dataset. In this work, we consider a high-dimensional setting where $p>N$.

Learning a linear decision rule to predict the target outcome $Y_0$ using covariate vector $\bX$ entails a classification problem. Empirical risk minimization (ERM) is often used to learn such a linear decision rule. Specifically, ERM often minimizes a convex surrogate of the loss function, i.e.,
\begin{equation}\label{eq:sample_loss}
\min_{\bm\theta_0} \ell(\bm\theta_0):=\mathbb{E}\left[ \phi\left\{Y_0(\bm X^\top\bm\beta_0+c_0)\right\}\mid R=0\right],
\end{equation}
where $\phi(\cdot)$ is a surrogate loss, and $\bm\theta_0=(\bm\beta_0, c_0)^\top$. By solving optimization problem ~\eqref{eq:sample_loss}, the decision rule, $d_{0}^*(\bm X)$, with the form $d_{0}^*(\bm X)=\mathrm{sgn}\left\{\bm X^\top\bm\beta_{0}^*+c_{0}^*\right\}$, can be used for prediction purposes, where $\bm\theta^*_0=(\bm\beta^*_0, c^*_0)^\top$ is the minimizer of optimization problem ~\eqref{eq:sample_loss}. Our goal is to use the auxiliary labels to improve the estimation of $\bm\theta^*_0$.

\subsection{Step one: learn a linear representation using MTL}

In this section, we introduce our proposed method, which consists of two steps. The first step is to learn a linear representation using MTL incorporating the auxiliary outcomes. Denote the index set of auxiliary outcomes as $\mathcal{J}=\left\{1,2,\cdots, J\right\}$.

In this work, we consider the following MTL method. We obtain a linear representation $\widehat{\bm w}_{\mathcal{J}}$ by solving 
\begin{equation}\label{eq:pooled_loss}
\min_{\bm w, \left\{c_j\right\}_{j\in \mathcal{J}}}\widehat{\mathbb{E}}_{N}\left[\sum_{j\in \mathcal{J}} \phi\left\{Y_j(\bm X^\top\bm w+c_j)\right\}\right]+\lambda_N\|\bm w\|_1,
\end{equation}
where $\lambda_N$ is a tuning parameter and $\widehat{\mathbb{E}}_N[\cdot]$ is the empirical expectation of both the target and source-only datasets. In this procedure, we estimate $J$ decision rules for $\{Y_j\}_{j\in \mathcal{J}}$, simultaneously. These decision rules are structured to learn a common parameter $\bm w$, which is the direction shared by all outcomes. In addition, the intercept represented by $c_j$'s can be different for each outcome to accommodate possible heterogeneity. Leveraging information from auxiliary outcomes (and/or the source-only dataset), the estimator $\widehat{\bm w}_{\mathcal{J}}$ can approach ${\bm w}^*_{\mathcal{J}}$ with a low estimation error, where ${\bm w}^*_{\mathcal{J}}$ is the minimizer of
\begin{equation*}
\min_{\bm w, \left\{c_j\right\}_{j\in \mathcal{J}}}\mathbb{E}\left[\sum_{j\in \mathcal{J}} \phi\left\{Y_j(\bm X^\top\bm w+c_j)\right\}\right].
\end{equation*}
Although the first step takes advantage of shared information across multiple outcomes, the estimator $\widehat{\bm w}_{\mathcal{J}}$ may be biased w.r.t $\bm\beta_{0}^*$, especially when ${\bm w}^*_{\mathcal{J}}$ is biased w.r.t $\bm\beta_{0}^*$.

\begin{remark}
In MTL~\eqref{eq:pooled_loss}, we only specify different intercepts to accommodate possible heterogeneity. Note that, we can allow any low-dimensional sub-vector of the coefficients to be different to accommodate heterogeneous effects. In this case, our theoretical results in Section~\ref{sec:theory} are still valid. For simplicity, in the main text, we focus on MTL~\eqref{eq:pooled_loss}.
\end{remark}

\subsection{Step two: a novel calibration step}

In this section, we present how to de-bias $\widehat{\bm w}_{\mathcal{J}}$ and construct an improved estimator for $\bm\beta_0^*$ through a novel calibration step. To start with, we decompose the bias of ${\bm w}^*_{\mathcal{J}}$ as the following,
\begin{eqnarray*}
\mathrm{bias}({\bm w}^*_{\mathcal{J}}):={\bm w}^*_{\mathcal{J}}-{\bm\beta}^*_0=(1-\gamma){\bm w}^*_{\mathcal{J}}-\bm \delta,
\end{eqnarray*}
where $\bm \delta:= {\bm\beta}^*_0-\gamma{\bm w}^*_{\mathcal{J}}$, i.e., ${\bm\beta}^*_0=\gamma{\bm w}^*_{\mathcal{J}}+\bm \delta$. The first term in this decomposition, $(1-\gamma){\bm w}^*_{\mathcal{J}}$, is along the direction of ${\bm w}^*_{\mathcal{J}}$, and thus, we refer to it as the \textsl{within-subspace bias}; the remaining term $\bm\delta$  is referred to as the \textsl{against-subspace bias}. To remove the bias in ${\bm w}^*_{\mathcal{J}}$, we need to adjust for both within-subspace and against-subspace biases.
Subsequently, we consider the following optimization problem,
\begin{equation}\label{eq:calibrated_loss_population}
\min_{\bm\delta, \gamma, c_0}\mathbb{E}\left[ \phi\left\{Y_0(\bm X^\top\bm\delta+\gamma \bm X^\top{\bm w}^*_{\mathcal{J}}+c_0)\right\}\right].
\end{equation}
The loss function in~\eqref{eq:calibrated_loss_population} incorporates two adjustments to ${\bm w}^*_{\mathcal{J}}$, which corresponds to the within-subspace bias and against-subspace bias. First, we calibrate the scaling parameter ${\gamma}$ along the subspace generated by ${\bm w}^*_{\mathcal{J}}$. This calibration eliminates the within-subspace bias. For instance, if ${\bm w}^*_{\mathcal{J}}=2\bm\beta_0^*$, then, setting $\gamma=1/2$ can eliminate such a bias.  Second, we calibrate the subspace generated by ${\bm w}^*_{\mathcal{J}}$ using ${\bm\delta}$. This calibration accounts for the against-subspace bias. If ${\bm w}^*_{\mathcal{J}}=\bm\beta_0^*-\bm e$, then setting $\bm\delta=\bm e$ can account for such a bias, where $\bm e=(1,0,\cdots, 0)^\top$. 

The decomposition of $\mathrm{bias}({\bm w}^*_{\mathcal{J}})$, i.e., $\mathrm{bias}({\bm w}^*_{\mathcal{J}})=(1-\gamma){\bm w}^*_{\mathcal{J}}-\bm \delta$ provides multiple options to adjust for possible bias. For each $\gamma$, we can obtain a different decomposition of the bias, and a different way to adjust for the bias. For example, when $\gamma=1$, the corresponding $\bm\delta={\bm\beta}^*_0-{\bm w}^*_{\mathcal{J}}$; when $\gamma=1/2$, the corresponding $\bm\delta={\bm\beta}^*_0-{\bm w}^*_{\mathcal{J}}/2$. Both $\gamma=1$ and $\gamma=1/2$ lead to a specification of $\bm\delta$ such that ${\bm\beta}^*_0=\gamma{\bm w}^*_{\mathcal{J}}+\bm \delta$. However, under different choices of $\gamma$, the $\bm\delta$'s may be different in terms of their $l_0$ and $l_1$ norms, resulting in different levels of difficulties in estimating them. For example, the contrast, ${\bm w}^*_{\mathcal{J}}-{\bm\beta}^*_0$, may not be sparse in $l_0$ norm nor small in $l_1$ norm. In this case, the contrast ${\bm w}^*_{\mathcal{J}}-{\bm\beta}^*_0$ may not be easy to estimate. Among all possible decompositions, the $\gamma$'s that can lead to a sparse ($l_0$ norm) or a small ($l_1$ norm) against-subspace bias, $\bm\delta$, are preferable. For ease of exposition, we focus on the $\bm\delta$ with the least $l_1$ norm. The results under $l_0$ norm can be found in the Online Supporting Information.


Denote the set of $\bm\delta$'s with the least $l_1$ norm as $\bm\delta^*$. To target the  $\gamma$ such that $\bm\delta\in \bm\delta^*$, we propose a special treatment as follows: we first separate the space of $\bm \delta$ into several domains such that in each domain, the solution is unique, i.e., $\gamma$ and $\bm\delta$ minimizing~\eqref{eq:calibrated_loss_population} or its penalized empirical version is unique; then, we select the final estimator through validation using a pre-split dataset. Below we introduce how these domains are defined, and show that, at least one solution with $\bm\delta\in \bm\delta^*$ must belong to one of these domains. 

\begin{remark}
This special treatment is not required if we only focus on the $\bm\delta$ with the least $l_1$ norm. We can directly solve~\eqref{eq:calibrated_loss_population} with a lasso penalty, if the $\bm\delta$ with the least $l_1$ norm is the targeted solution. The proposed procedure provides a unified procedure with theoretical guarantees when either the $\bm\delta$ with the least $l_1$ norm or the $\bm\delta$ with the least $l_0$ norm leads to the optimal estimation of $\bm\beta^*_0$.
\end{remark}

We construct the following domains $$\Gamma_k=\left\{\bm \delta=(\delta_1, \delta_2,\cdots, \delta_p)^\top: \delta_k=0\right\}.$$ Due to the strict convexity of $\phi$ and the assumption that the coordinates of $\bm X$ are not linearly dependent,  for any $k\in S_{\mathcal{J}}^*$, there exists a unique $\gamma$ such that $\bm\beta_{0}^*-\gamma {\bm w}^*_{\mathcal{J}}\in\Gamma_k$, where $S_{\mathcal{J}}^*$ is the set of indexes of the non-zero coefficients of ${\bm w}^*_{\mathcal{J}}$. This implies that the objective function in ~\eqref{eq:calibrated_loss_population} on each $\Gamma_k$ has a unique minimizer, for any $k\in S_{\mathcal{J}}^*$.

Lemma~\ref{lemma:1} further implies that to determine the $\gamma$ such that $\bm\delta\in \bm\delta^*$, we only need to solve the optimization problem~\eqref{eq:calibrated_loss_population} within each domain, and there exists a targeted solution, i.e., the $\gamma$ and the associated $\bm\delta$, such that the associated $\bm\delta$ belongs to one of these domains and $\bm\delta\in \bm\delta^*$. Lemma~\ref{lemma:1} also holds if $\bm\delta^*$ is defined as the set of $\bm\delta$'s with the least $l_0$ norm.
\begin{lemma}\label{lemma:1}
There exists a minimizer of the optimization problem~\eqref{eq:calibrated_loss_population}, $\gamma$ and $\bm\delta$, satisfying that $\bm\delta\in \cup_{k\in S_{\mathcal{J}}^*}{\Gamma}_k$ and $\bm\delta\in \bm\delta^*$.
\end{lemma}

Motivated by this, given an index set $S$, we consider a set of optimization problems
\begin{equation}\label{eq:calibrated_loss_constraint}
\min_{\bm\delta\in {\Gamma}_k, \gamma, c_0}\widehat{\mathbb{E}}_n\left[ \phi\left\{Y_0(\bm X^\top\bm\delta+\gamma \bm X^\top\widehat{\bm w}_{\mathcal{J}}+c_0)\right\}\right]+\widetilde{\lambda}_n\|\bm\delta\|_1
\end{equation}
for $k\in S$, where $\widetilde{\lambda}_n$ is a tuning parameter. The optimization problem~\eqref{eq:calibrated_loss_constraint} contains the empirical version of the objective function in~\eqref{eq:calibrated_loss_population}, but constrains the domain of $\bm\delta\in {\Gamma}_k$. In each $\Gamma_k$, the solution of optimization~\eqref{eq:calibrated_loss_constraint} is unique for $k\in {S}$; when $S^*_{\mathcal{J}}\subset S$, the optimization is guaranteed to identify $\bm\delta$, which has the least $l_1$ norm. 


To select the final estimator among the proposed domains with different $k$'s, we propose a cross-fitting procedure. First, we calculate the minimizer of 
\begin{equation*}
\min_{\bm\delta\in {\Gamma}_k, \gamma, c_0}\widehat{\mathbb{E}}_n\left[ \phi\left\{Y_0(\bm X^\top\bm\delta+\gamma \bm X^\top\widehat{\bm w}_{\mathcal{J}}+c_0)\right\}\right]+\widetilde{\lambda}_n\|\bm\delta\|_1,
\end{equation*}
for each $k\in S$, denoted as $\widehat{\bm\delta}(k)$, $\widehat{\gamma}(k)$, $\widehat{c}(k)$. Subsequently, we have $\widehat{\bm\beta}(k)=\widehat{\bm\delta}(k)+\widehat{\gamma}(k)\widehat{\bm w}_{\mathcal{J}}$. 
Then, we pick one of these estimates based on the loss evaluated on a pre-split sample as the final estimate. We denote the selected estimate as $\widehat{\bm\beta}_0$ and $\widehat{c}_0$. The entire procedure is illustrated as Algorithm \ref{algo:1}.

\begin{algorithm}
\small
	\caption{Estimation of the optimal decision rule using auxiliary outcomes \label{algo:1}}
    	\SetAlgoLined
    	\KwIn{$\left\{(\bX_i, Y_{0,i}, Y_{1,i}, \cdots, Y_{J,i})\right\}_{i=1}^n$ and $\left\{(\bX_i, Y_{1,i}, \cdots, Y_{J,i})\right\}_{i=n+1}^N$.}
    	\KwOut{ A decision rule $\widehat{d}(X)=\mathrm{sgn}\left\{\bm X^\top\widehat{\bm\beta}_0+\widehat{c}_0\right\}$.}
    \label{algo1:1} Randomly split the samples $\mathcal{I}_1$ and $\mathcal{I}_2$ with an equal size\;
		\label{algo1:2} Using samples in $\mathcal{I}_1$, we obtain the estimation $\widehat{\bm w}_{\rm pool, \mathcal{I}_1}$ by solving 
\begin{equation*}
\min_{\bm w, \left\{c_j\right\}_{j=0}^J}\widehat{\mathbb{E}}_{N,\mathcal{I}_1}\left[\sum_{j\in \mathcal{J}} \phi\left\{Y_j(\bm X^\top\bm w+c_j)\right\}\right]+\lambda_N\|\bm w\|_1,
\end{equation*}
where $\widehat{\mathbb{E}}_{N,\mathcal{I}_1}[\cdot]$ is the empirical expectation on $\mathcal{I}_1$ and $\lambda_N$ is tuned by cross-validation\;
		\label{algo1:3} Using samples in $\mathcal{I}_1$, we obtain the calibrate estimator $\widehat{\bm\beta}_{\rm cal, \mathcal{I}_1}(k)$ and $\widehat{c}_{ \mathcal{I}_1}(k)$ for each $k\in \widehat{S}_{\rm pool, \mathcal{I}_1}$ by solving
\begin{equation*}
\min_{\bm\delta\in \Gamma_k, \gamma, c_0}\widehat{\mathbb{E}}_{n, \mathcal{I}_1}\left[ \phi\left\{Y_0(\bm X^\top\bm\delta+\gamma \bm X^\top\widehat{\bm w}_{\rm pool, \mathcal{I}_1}+c_0)\right\}\right]+\widetilde{\lambda}_{n,1}\|\bm\delta\|_1,
\end{equation*}
where $\widehat{S}_{\rm pool, \mathcal{I}_1}$ is the index of the non-zero coefficients in $\widehat{\bm w}_{\rm pool, \mathcal{I}_1}$, and $\widetilde{\lambda}_{n,1}$ is tuned by cross-validation\;
		
		\label{algo1:4} Choose $\widehat{\bm\beta}_{\rm cal, \mathcal{I}_1}=\widehat{\bm\beta}_{\rm cal, \mathcal{I}_1}(k^*)$ and $\widehat{c}_{ \mathcal{I}_1}=\widehat{c}_{ \mathcal{I}_1}(k^*)$, where $k^*$ solves
\begin{equation*}
\min_k \widehat{\mathbb{E}}_{n, \mathcal{I}_2}\left[\phi\left\{Y_0(\bm X^\top\widehat{\bm\beta}_{\rm cal, \mathcal{I}_1}(k)+\widehat{c}_{ \mathcal{I}_1}(k))\right\}\right],
\end{equation*}
where $\widehat{\mathbb{E}}_{n,\mathcal{I}_2}[\cdot]$ is the empirical expectation on $\mathcal{I}_2$ with $R_i=1$\;
	
		\label{algo1:5} Replace $\mathcal{I}_1$ by $\mathcal{I}_2$ and $\mathcal{I}_2$ by $\mathcal{I}_1$, and repeat Steps~2 -~4. Obtain $\widehat{\bm\beta}_{0}$ and $\widehat{c}_0$ by
		\begin{equation*}
			\widehat{\bm\beta}_0=\left(\widehat{\bm\beta}_{\rm cal, \mathcal{I}_1}+\widehat{\bm\beta}_{\rm cal, \mathcal{I}_2}\right)/2,
		\end{equation*}
and
		\begin{equation*}
			\widehat{c}_0=\left(\widehat{c}_{\rm cal, \mathcal{I}_1}+\widehat{c}_{\rm cal, \mathcal{I}_2}\right)/2.
		\end{equation*}
\end{algorithm}

\section{Theoretical properties}
\label{sec:theory}

To provide theoretical support for the effectiveness and applicability of the proposed method, we investigate the convergence rate of the proposed estimator. The proof of all the lemmas, theorems and corollaries can be found in the Online Supporting Information.
First, we provide the convergence rate of $\widehat{\bm w}_{\mathcal{J}}$ under the following assumptions.
\begin{assumption}
\label{cond:1} There is a constant $R $ such that $\|\bm X\|_{\infty}$, $\sup_{\bm X}|\bm X^\top\bm w^*_{\mathcal{J}}|$, and $|c^*_{j}|$'s are upper bounded by $R$ with probability $1$.
\end{assumption}
\begin{assumption}
\label{cond:2} Define $\widetilde{\bm X}=(1, \bm X)$. There is a constant $\lambda_{\min}$ such that the smallest eigenvalue of $\mathbb{E}[\widetilde{\bm X}\widetilde{\bm X}^\top]$ is lower bounded by $\lambda_{\min}$.
\end{assumption}

We denote the index set of the non-zero coordinates of $\bm\beta^*_0$ as $S^*$, and the index set of the non-zero coordinates of $\bm w^*_{\mathcal{J}}$ as $S^*_{\mathcal{J}}$. The cardinality of $S^*$ is denoted as $s^*$, and the cardinality of $S^*_{\mathcal{J}}$ is denoted as $s^*_{\mathcal{J}}$. To characterize the commonalities and differences across auxiliary outcomes, we also define $\alpha$ and $h_b$ such that 
\begin{eqnarray*}
\mathrm{var}\left(\sum_{j\in \mathcal{J}}Y_j\phi'\left\{ Y_j(\bm X^\top\bm w^*_{\mathcal{J}}+c_j^*)\right\}X_k\mid \bm X\right)&\leq& CJ^{\alpha},\\
\sup_{j\in \mathcal{J}}\left|E\left(Y_j\phi'\left\{ Y_j(\bm X^\top\bm w^*_{\mathcal{J}}+c_j^*)\right\}X_k\mid \bm X\right)\right|&\leq& Ch_b.
\end{eqnarray*}
Without any additional assumptions, we can take $\alpha=2$ and $h_b=O(1)$. Under certain additional assumptions, $\alpha$ and $h_b$ can be chosen differently. For example, when $Y_j$'s are mutually independent conditional on $\bm X$, we can take $\alpha=1$; when 
\begin{eqnarray}\label{eq:model_assumption_h_b}
P(Y_j=1\mid \bm X)/\left\{1-P(Y_j=1\mid \bm X)\right\}=\phi'\left\{-\bm X^\top\bm w^*_{\mathcal{J}}-c_j^*\right\}/\phi'\left\{\bm X^\top\bm w^*_{\mathcal{J}}+c_j^*\right\},
\end{eqnarray}
we can take $h_b=0$. Note that, when $\phi(\cdot)$ is a logistic loss, the model assumption~\eqref{eq:model_assumption_h_b} is equivalent to logistic model assumptions with the same coefficients and different intercepts for auxiliary outcomes. From these examples, we can see that $\alpha$ controls the mutual dependence between $Y_j$'s conditional on $\bm X$, and $h_b$ controls the bias of $P(Y_j=1\mid \bm X)$ w.r.t. the model $P(Y_j=1\mid \bm X)/\left\{1-P(Y_j=1\mid \bm X)\right\}=\phi'\left\{-\bm X^\top\bm w^*_{\mathcal{J}}-c_j^*\right\}/\phi'\left\{\bm X^\top\bm w^*_{\mathcal{J}}+c_j^*\right\}$. Thus, by incorporating $\alpha$ and $h_b$, our theoretical results can accommodate $Y_j$'s that are dependent, and model mis-specifications w.r.t the model in~\eqref{eq:model_assumption_h_b}. 

Furthermore, to characterize the difference between auxiliary outcomes and the target outcome, we define $h$ such that
\begin{eqnarray*}
\inf_{\lambda}\sup_{j\in \mathcal{J}} \left\|\lambda\bm w_j^*-\bm\beta_0^*\right\|_1\leq h,
\end{eqnarray*}
where $\bm w_j^*$ is the solution to each single task, i.e.,
$
    \min_{\bm w,c_j}\mathbb{E}\left[\phi\left\{Y_j(\bm X^\top\bm w+c_j)\right\}\right].
$

\begin{remark}
The definition
\begin{eqnarray*}
\inf_{\lambda}\sup_{j\in \mathcal{J}} \left\|\lambda\bm w_j^*-\bm\beta_0^*\right\|_1\leq h
\end{eqnarray*}
enables a bound for the cosine-angle between $\bm w_j^*$ and $\bm\beta_0^*$ (decision boundaries differ up to an intercept). Specifically, for any $\lambda\not = 0$, we have
\begin{eqnarray*}
\sup_{j\in \mathcal{J}}\frac{\left|\left(\bm\beta_0^*\right)^\top \bm w_j^*\right|}{\left\|\bm\beta_0^*\right\|_2\left\|\bm w_j^*\right\|_2}=\sup_{j\in \mathcal{J}}\frac{\left|\left(\bm\beta_0^*\right)^\top \lambda\bm w_j^*\right|}{\left\|\bm\beta_0^*\right\|_2\left\|\lambda\bm w_j^*\right\|_2}
\geq \frac{\left\|\bm\beta_0^*\right\|_2-\sup_{j\in \mathcal{J}}\left|\lambda\bm w_j^*-\bm\beta^*_0\right|_1}{\left\|\bm\beta_0^*\right\|_2+\sup_{j\in \mathcal{J}}\left|\lambda\bm w_j^*-\bm\beta^*_0\right|_1}.
\end{eqnarray*}
Thus, under $\inf_{\lambda}\sup_{j\in \mathcal{J}} \left\|\lambda\bm w_j^*-\bm\beta_0^*\right\|_1\leq h$, we have
\begin{eqnarray*}
\sup_{j\in \mathcal{J}}\frac{\left|\left(\bm\beta_0^*\right)^\top \bm w_j^*\right|}{\left\|\bm\beta_0^*\right\|_2\left\|\bm w_j^*\right\|_2}\geq \frac{\left\|\bm\beta_0^*\right\|_2-h}{\left\|\bm\beta_0^*\right\|_2+h}.
\end{eqnarray*}
\end{remark}

Given these notations, we have the following lemma on the convergence rate of $\widehat{\bm w}_{\mathcal{J}}$.
\begin{lemma}\label{thm:pooled_convergence}
Under the Conditions~\ref{cond:1} and~\ref{cond:2}, with probability approaching to $1$, we have
\begin{eqnarray*}
\left\|\widehat{\bm w}_{\mathcal{J}}-\bm w^*_{\mathcal{J}}\right\|_1&=&O_p\left(\frac{s^*\lambda_N}{\left|\mathcal{J}\right|}+C_{\Sigma}h\right),\\
\left\|\widehat{\bm w}_{\mathcal{J}}-\bm w^*_{\mathcal{J}}\right\|_2^2&=&O_p\left(\left\{\frac{s^*\lambda_N^2}{\left|\mathcal{J}\right|^2}+\frac{\lambda_N C_{\Sigma}h}{\left|\mathcal{J}\right|}\right\}\wedge \left(C_{\Sigma}h\right)^2\right),
\end{eqnarray*}
if $\lambda_N\gg \sqrt{\frac{\log J}{N}}$ and $\lambda_N\geq \sqrt{\frac{2C(J^{\alpha}+J^2h_b^2)\log p}{N}}\vee \frac{J\log p}{N}$, where $C_{\Sigma}$ can be found in the Online Supporting Information.
\end{lemma}

Compared with the existing literature, the convergence rate of $\widehat{\bm w}_{\mathcal{J}}$ could be faster than those in \cite{li2020transfer}. Specifically, when $J\to +\infty$, $\alpha<2$, and $h_b=0$, the convergence rate of $\widehat{\bm w}_{\mathcal{J}}$ is faster than those in \cite{li2020transfer}, when $\lambda_N\asymp \sqrt{\frac{J^{\alpha}\log p}{N}}$. In practice, we can choose $\lambda_N$ using cross-validation.

To investigate the theoretical property of $\widehat{\bm\beta}_0$, we further introduce the following assumptions.
\begin{assumption}\label{cond:4} 
We assume that $\sup_{\bm X}|\bm X^\top\bm\beta_0^*|\leq R$, and $|c_0^*|\leq R$. We also require that the minimizer of $$\min_{\gamma}\|\bm\beta_{0}^*-\gamma {\bm w}^*_{\mathcal{J}}\|_0\text{ s.t. } \bm\beta_{0}^*-\gamma {\bm w}^*_{\mathcal{J}}\in\cup_{k\in S} \Gamma_k,$$ is bounded by $R$. We assume that $\frac{s^*\log p}{n}\to 0$.
\end{assumption}
\begin{assumption}\label{cond:6} 
Define $\widetilde{\bm X}_{j}=(\bm X^\top\bm w^*_{\mathcal{J}}, \bm X_{-j}, 1)$, where $\bm X_{-j}$ is the vector of covariates $\bm X$ excluding the $j$th covariate. We assume that there is a constant $\widetilde{\lambda}_{\min}$ such that the smallest eigenvalue of $\mathbb{E}[\widetilde{\bm X}_{j}\widetilde{\bm X}_{j}^\top]$ is lower bounded by $\widetilde{\lambda}_{\min}$ for all $j\in S$.
\end{assumption}

Assumption~\ref{cond:4} assumes a uniform bound on the design matrix for technical simplicity. 
Assumption~\ref{cond:6} assumes a uniform lower bound for the eigenvalues of the design matrix for all $j\in S$. 

\begin{theorem}\label{thm:rate_beta}
	Under Assumptions~\ref{cond:1} -~\ref{cond:6}, taking $\lambda_N\gg \sqrt{\frac{\log J}{N}}$, $\lambda_N\geq \sqrt{\frac{2C(J^{\alpha}+J^2h_b^2)\log p}{N}}\vee \frac{J\log p}{N}$, and $\widetilde{\lambda}_n\asymp \sqrt{\frac{\log p}{n}}$, we have 
	\begin{eqnarray*}
    && \max\left\{\left\|\widehat{\bm\beta}_0-\bm\beta^*_0\right\|_2^2,\left\|\widehat{c}_0-c^*_0\right\|_2^2\right\}\\
    &\lesssim& \left\{\frac{s^*\lambda_N^2}{\left|\mathcal{J}\right|^2}+\frac{\lambda_N C_{\Sigma}h}{\left|\mathcal{J}\right|}\right\}\wedge \left(C_{\Sigma}h\right)^2+\left\{\widetilde{\lambda}_n^2+\widetilde{\lambda}_n h^*_{\bm\delta}\right\}+\frac{\log |S|\vee n}{n}
     \end{eqnarray*}
	with probability approaching to $1$, where $h^*_{\bm\delta}$ is the minimizer of $\min_{\gamma}\|\bm\beta_{0}^*-\gamma {\bm w}^*_{\mathcal{J}}\|_1\text{ s.t. } \bm\beta_{0}^*-\gamma {\bm w}^*_{\mathcal{J}}\in\cup_{k\in S} \Gamma_k$. 
\end{theorem}

The resultant rate in Theorem~\ref{thm:rate_beta} is structured as the sum of three terms. The first term, $\left\{\frac{s^*\lambda_N^2}{\left|\mathcal{J}\right|^2}+\frac{\lambda_N C_{\Sigma}h}{\left|\mathcal{J}\right|}\right\}\wedge \left(C_{\Sigma}h\right)^2$, is related to the estimation error of $\widehat{\bm w}_{\mathcal{J}}$. The second term, $\widetilde{\lambda}_n^2+\widetilde{\lambda}_n h^*_{\bm\delta}$, is associated with the minimal against-subspace bias of ${\bm w}^*_{\mathcal{J}}$. The third term, $\frac{\log |S|\vee n}{n}$, accounts for the variability of selecting $k^*$ in Step~4 of Algorithm~\ref{algo:1}. 

Compared with the convergence rate of using only $n$ samples and the target label $Y_0$, i.e., $O_p\left(s^*\frac{\log p}{n}\right)$, the convergence rate shown in Theorem~\ref{thm:rate_beta} can be faster. For example, when $N\gg n$ and $C_{\Sigma}h\ll s^*\sqrt{\frac{\log p}{n}}\sqrt{\frac{N}{n}}$, the first term is faster than $s^*\frac{\log p}{n}$. For the second term, when $h^*_{\bm \delta}\ll s^*\sqrt{\frac{\log p}{n}}$, we have $\widetilde{\lambda}_n^2+\widetilde{\lambda}_n h^*_{\bm\delta}\ll s^*\frac{\log p}{n}$. The third term is always negligible compared with $s^*\frac{\log p}{n}$. Hence, when $N$ is sufficiently large compared with $n$, $h^*_{\bm \delta}\ll s^*\sqrt{\frac{\log p}{n}}$ can lead to a  convergence rate faster than $O_p\left(s^*\frac{\log p}{n}\right)$.

\begin{remark}
Different from \cite{li2020transfer}, where a necessary condition for a convergence rate faster than $O_p\left(s^*\frac{\log p}{n}\right)$ is $N\gg n$, Theorem~\ref{thm:rate_beta} shows that, when $N=n$, it is also possible to have a  convergence rate faster than $O_p\left(s^*\frac{\log p}{n}\right)$. For example, when  $(|\mathcal{J}|^{\alpha-2}+h_b^2)\to 0$ with $C_{\Sigma}h\sqrt{|\mathcal{J}|^{\alpha-2}+h_b^2}\ll s^*\sqrt{\frac{\log p}{n}}$ and $h^*_{\bm \delta}\ll s^*\sqrt{\frac{\log p}{n}}$,  the convergence rate is faster than $O_p\left(s^*\frac{\log p}{n}\right)$. As a necessary condition for a convergence rate faster than $O_p\left(s^*\frac{\log p}{n}\right)$ when $N=n$, we require that $|\mathcal{J}|^{\alpha-2}+h_b^2\to 0$. This requirement holds when we have many weakly dependent auxiliary outcomes (i.e., $\alpha<2$) with a small bias (i.e., $h_b\to 0$) against the model in \eqref{eq:model_assumption_h_b}.
\end{remark}

\begin{remark}
To achieve the requirement that $C_{\Sigma}h\ll s^*\sqrt{\frac{\log p}{n}}\sqrt{\frac{N}{n}}$ and $h^*_{\bm \delta}\ll s^*\sqrt{\frac{\log p}{n}}$, the choice of $\mathcal{J}$ is important. Without an appropriate selection of $\mathcal{J}$, the convergence rate $\widehat{\bm\beta}_0$ is not necessarily faster than the convergence rate of using only  the target label $Y_0$ (with sample size $n$); this phenomenon is referred to as the negative transfer \citep{tian2022transfer}. The transferable source detection algorithm proposed in \cite{tian2022transfer} can also be applied in our proposed method to avoid a possible negative transfer.
\end{remark}

\begin{remark}
Lemma~\ref{thm:pooled_convergence} and Theorem~\ref{thm:rate_beta} are stated under the definition that
\begin{eqnarray*}
\inf_{\lambda}\sup_{j\in \mathcal{J}} \left\|\lambda\bm w_j^*-\bm\beta_0^*\right\|_1\leq h.
\end{eqnarray*}
The definition of $h$ depends on the $l_1$ norm of $\lambda\bm w_j^*-\bm\beta_0^*$. In the Online Supporting Information, we also derive another convergence rate using the $l_0$ norm of $\lambda\bm w_j^*-\bm\beta_0^*$, i.e.,
\begin{eqnarray*}
\inf_{\lambda}\sup_{j\in \mathcal{J}} \left\|\lambda\bm w_j^*-\bm\beta_0^*\right\|_0.
\end{eqnarray*}
\end{remark}

\section{Simulations}
\label{sec:sim}

In this section, we conduct simulations to compare the performance of our proposed method with other existing approaches (e.g., MTL approaches and other transfer learning approaches). One of the comparison methods, referred to as the baseline approach, uses solely the target outcome. For the baseline approach, we directly solve
\begin{equation*}
\min_{\bm\beta_0, c_0}\widehat{\mathbb{E}}_n\left[ \phi\left\{Y_0(\bm X^\top\bm\beta_0+c_0)\right\}\right]+\lambda_n\|\bm\beta_0\|_1,
\end{equation*}
where the logistic loss is chosen for $\phi(\cdot)$  and $\lambda_n$ is tuned by cross-validation. The other approaches for comparison include a direct transfer learning approach and two MTL approaches. The direct transfer learning approach implements a modified Algorithm~\ref{algo:1}, where one fixes $\gamma=1$, and $c_j$'s in Step one are assumed to be the same. This modified algorithm can be considered an extension of the oracle Trans-Lasso Algorithm (TransferDirect) proposed in \cite{li2020transfer}. The multi-task learning approach 1 (MultiTask1) extends the algorithm proposed in \cite{obozinski2008high} using a logistic loss with a grouped lasso penalty. The multi-task learning approach 2 (MultiTask2) is the approach minimizing
\begin{equation*}
\min_{\bm w, \left\{c_j\right\}_{j\in \mathcal{J}\cup 0}}\widehat{\mathbb{E}}_{n}\left[\sum_{j\in \mathcal{J}} \phi\left\{Y_j(\bm X^\top\bm \beta+c_j)\right\}\right]+\lambda_N\|\bm \beta\|_1.
\end{equation*}
MultiTask2 shares a similar loss function as the MTL used in Step one for our proposed approach. Comparing the proposed method with the baseline approach, we can examine the performance gained from using the auxiliary outcome. Comparing the proposed method with the direct transfer learning, we can see the benefit of the proposed method over the existing transfer learning approaches due to the capability to accommodate heterogeneous models in the MTL step and the novel calibration step. By comparison to the two MTL approaches, we can examine the difference between the transfer learning and MTL approaches when focusing on the problem of target label prediction.

Let $\bm\beta_{U_0}$ be the coefficients related to the latent variable $U_0$ for the target outcome. We generate experimental data following the simulation scenarios below:
\begin{enumerate}[label=(\Roman*)]
	\item \label{sim:1} We set $n=N$. Let $\bm\beta_{U_0}=(1, -1,1,-1, 0,\cdots,0, 0.5,-0.5,2,-2,0.5,0.5,0, \cdots,0)^\top$ and $U_0=5G(\bm X^\top\bm\beta_{U_0})+0.2\epsilon_{U_0}$, where $\epsilon_{U_0}$ follows a standard normal distribution. The function $G(\cdot)$ is the cumulative distribution function of a standard normal distribution. Set $\widetilde{U}=5G(\bm X^\top\bm\beta_{\widetilde{U}})+0.2\epsilon_{\tilde{U}}$, where $\epsilon_{\widetilde{U}}$ follows a standard normal distribution, and the $q$-th coordinate of $\bm\beta_{\widetilde{U},q}$ satisfies that $\bm\beta_{\widetilde{U},q}=\bm\beta_{U_0,q}$ for $q\not= 2,4$ and $\bm\beta_{\widetilde{U},2}=\bm\beta_{\widetilde{U},4}=1$. The target outcome is generated by setting $Y_0=\mathrm{sgn}\left\{U_0-u_{0,1/4}\right\}$, where $u_{0,1/4}$ is the first quartile of $U_0$. We further introduce a weighting parameter $\alpha$ and generate the auxiliary outcome $Y_1$ by setting $Y_1=\mathrm{sgn}\left\{U_1-u_{1,3/4}\right\}$, where $U_1=(1-\alpha)U_0+\alpha\widetilde{U}$, and $u_{1,3/4}$ is the third quartile of $U_1$.
	\item \label{sim:2} We set $n=0.2N$ and $\bm\beta_{U_0}=(1, -1,1,-1, 0,\cdots,0)^\top$. We generate $U_0$ based on a binomial distribution $B(8, G(\bm X^\top\bm\beta_{U_0}))$, where the number of trials equals $8$ and the success probability equals $G(\bm X^\top\bm\beta_{U_0})$. Then, we corrupt this $U_0$: when $U_0\leq 3$, we set $U_1=U_0+B(3, \alpha)$; when $U_0>4$, we set $U_1=U_0-B(3, \alpha)$. The target outcome is set as $Y_0=1\left\{U_0>0\right\}$; the auxiliary outcomes are set as $Y_j=1\left\{U_1-(2j-1)\right\}$, where $j=1,2,3,4$.
\end{enumerate}

For the choice of covariate vector $\bm X$, we have the following two designs. In Design I, the covariate vector $X$ follows Gaussian distribution $N(\bm 0, \bm I_p)$. In Design II, we first generate a $p$-dimensional vector following $N(\bm 0, \bm \Sigma_p)$, where the $(l,k)$-th coordinate of $\bm \Sigma_p$ is $0.5^{|l-k|}$; then, for $l=1,\cdots,\lfloor p/4\rfloor$, we replace the $4l$-th coordinates in the generated vector with a binary variable. This binary variable is $1$ if and only if the generated coordinate is greater than 0. Compared with Design I, Design II has correlated covariates, and the covariates include both discrete and continuous variables. We test our methods using both designs for Scenarios~\ref{sim:1} and~\ref{sim:2}. In addition, both Scenarios~\ref{sim:1} and~\ref{sim:2} involve a parameter $\alpha$. When $\alpha=0$, because $U_0=U_1$ in both settings, we can show that $\bm\beta_0^*=\gamma\bm w_{\mathcal{J}}^*$ for some $\gamma$. With the increase of $\alpha$, $\bm w_{\mathcal{J}}^*$ involves more against-subspace bias. 

To compare the performance of different approaches, we generate a testing dataset with sample size $n=10^4$ and calculate two scores. Let $\widehat{\mathbb{E}}_{\rm test}[\cdot]$ be the empirical expectation calculated using the testing dataset. The first score is the accuracy. Given an estimated decision rule $\widehat{d}_0(\bm X)=\mathrm{sgn}\left\{\bm X^\top\widehat{\bm\beta}_0+\widehat{c}_0\right\}$, the accuracy is defined as $\widehat{\mathbb{E}}_{\rm test}\left[1\left\{Y_0=\widehat{d}_0(\bm X)\right\}\right]$. The other score is the rank correlation. We calculate the rank correlation between $\bm X^\top\bm\beta_{U_0}$ and $\bm X^\top\widehat{\bm\beta}_0$ and use it as a proxy of the estimation error.

In these simulations, we vary the sample size of the training dataset from $N=200$, $350$, to $500$ and fix $p=1000$. In Scenario~\ref{sim:1}, we change $\alpha$ from $0$ to $1$ with an increment of 0.25. In Scenario~\ref{sim:2}, we change $\alpha$ from $0$ to $0.3$ with an increment of 0.1. We repeat each simulation setting for $500$ times. 

\begin{figure}[!hbt]
\centering
  \includegraphics[scale=0.7]{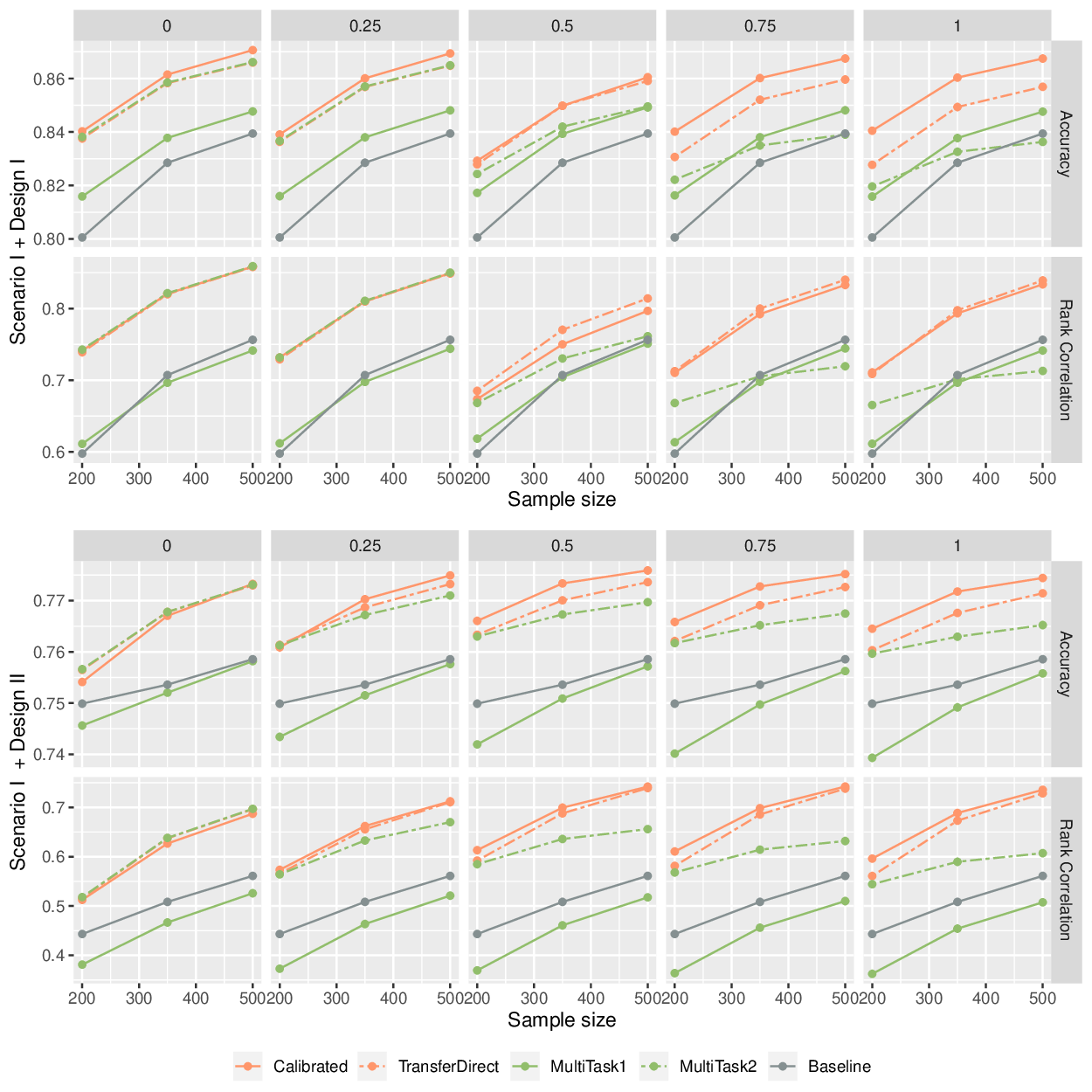}
  \caption{Simulation results for Scenario I with the change of sample size and $\alpha$.\label{fig:sim_score}}
\end{figure}

\begin{figure}[!hbt]
\centering
  \includegraphics[scale=0.7]{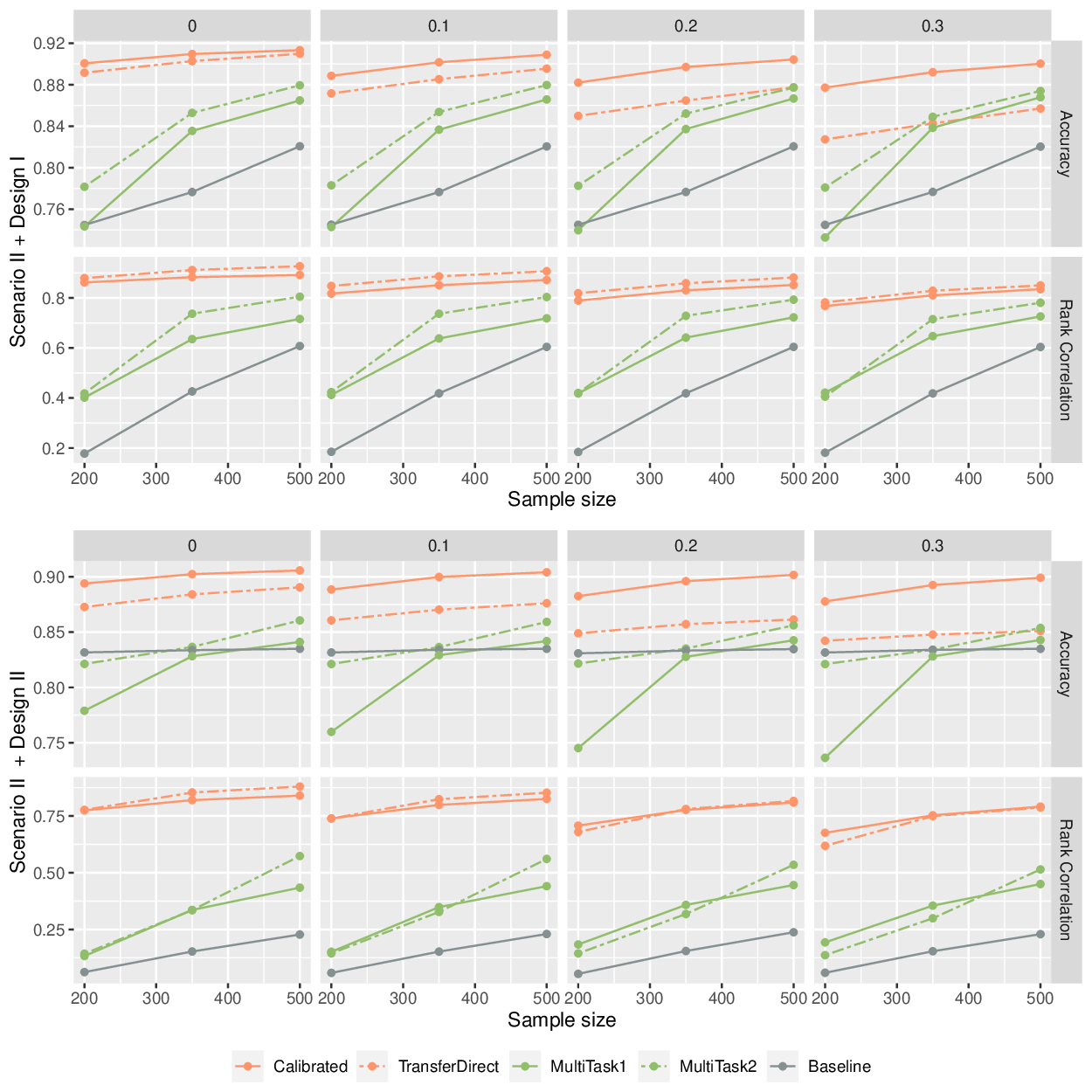}
  \caption{Simulation results for Scenario II with the change of sample size and $\alpha$.\label{fig:sim2_score}}
\end{figure}

Figures~\ref{fig:sim_score} and~\ref{fig:sim2_score} illustrate how the performance metrics vary with the increase of sample sizes and $\alpha$, for simulation Scenarios~\ref{sim:1} and~\ref{sim:2}, respectively. In Scenario~\ref{sim:1}, in terms of the accuracy and the rank correlation, the proposed method outperforms the baseline approach regardless of the change of sample sizes and $\alpha$. Compared with MultiTask1 and MultiTask2, our proposed method and TransferDirect are more robust w.r.t the change of $\alpha$; compared with TransferDirect, our proposed method shows great advantages in terms of prediction accuracy. In Scenario~\ref{sim:2}, our proposed method also performs better than other methods regardless of the change of sample sizes and $\alpha$.
 
\section{Application to predicting whether patients will achieve clinically significant difference after total hip arthroplasty}
\label{sec:real}

In this section, we apply our proposed method to predict the event of not achieving MCID for THA patients. The data was obtained from Patient-Reported Outcomes Measurement Information System (PROMIS)-10 at the University of Florida Health. In this dataset, we have 202 patients who underwent an index THA hospitalization, and we consider 13 variables, including race, Risk Assessment and Prediction Tool (RAPT), and preoperative HOOS JR survey responses, as covariates. The target outcome is chosen as the event of the (overall) improvement not achieving the anchor-based MCID \citep{fontana2019can}; $Y_0 = 1$, if the patient did not achieve the MCID and $Y_0 = -1$, otherwise. To obtain auxiliary outcomes, we use other outcomes (e.g., improvement in pain) on the survey questionnaire or their transformations as the auxiliary outcomes. For example, if an outcome is continuous or ordinal, we use the quartiles to split its value into four groups and then use the group indicators as the auxiliary outcomes.

To compare with different methods, we randomly split the dataset into a training dataset (70\% of the entire dataset) and a testing dataset (30\% of the entire dataset). We fit the proposed method and other comparison methods on the training dataset and calculate the accuracy and the Area Under Receiver Operating Characteristic Curve (AUC) on the testing dataset. The entire procedure is repeated 500 times. The mean and standard error of the accuracy and AUC are reported in Table~\ref{tab:real_data_result}. The results therein show that, the proposed method achieves the highest accuracy compared with all other methods in terms of prediction accuracy; the proposed method performs comparable to MultiTask2 in terms of the AUC. The Online Supporting Information provides an additional application to total knee arthroplasty patients. In this additional application, the auxiliary outcomes are chosen as the responses to the preoperative survey. Different from the application to THA patients, in this additional application, there are many patients who only responded to the preoperative survey, and thus they can be used as a source-only dataset. Through analysis in this additional application, we compare the performance of different methods under the setting where a large source-only dataset is available. 

\begin{table}
	\caption{Comparison of the mean (standard error) of accuracy and the Area Under Receiver Operating Characteristic Curve (AUC) estimated from five methods by repeated sample-splittings of the real data}
	\label{tab:real_data_result}
	\begin{center}
		\begin{tabular}{ccc}
			\toprule
			Method & Accuracy & AUC \\
			\hline\\[-2.5ex]
			Proposed & 0.746 (0.003) & 0.712 (0.004)\\
			TransferDirect & 0.740 (0.003) & 0.701 (0.004)\\
			MultiTask1 & 0.712 (0.003) & 0.660 (0.004)\\
			MultiTask2 & 0.739 (0.003) & 0.713 (0.004)\\
			Baseline & 0.731 (0.003) & 0.663 (0.004)\\
			[0.5ex]
			\bottomrule
		\end{tabular}
	\end{center}
\end{table}

\section{Discussion}
\label{sec:diss}

In this work, we develop a robust and flexible learning approach to improving high-dimensional linear decision rule estimation using the auxiliary outcomes. Our approach involves a two-step estimation procedure that takes advantage of the information provided by auxiliary outcomes and retains robustness against the bias introduced by auxiliary outcomes. Our novel bias decomposition allows for weaker required conditions and achieves superior performance against existing approaches. 

One possible extension is to propose a transfer learning approach under a more relaxed condition. As one of the major contributions, our proposed estimator can achieve faster convergence rates when $\inf_{\lambda}\sup_{j\in \mathcal{J}} \left\|\lambda\bm w_j^*-\bm\beta_0^*\right\|_1\ll s^*\sqrt{\log p/n}$, which is less restrictive than the requirement in \cite{li2020transfer, tian2022transfer}. A more mild condition is, for example, that $\sup_{j\in \mathcal{J}} \inf_{\lambda}\left\|\lambda\bm w_j^*-\bm\beta_0^*\right\|_1\ll s^*\sqrt{\log p/n}$. It could be interesting to see a transfer learning approach that achieves faster convergence rates under this more mild condition.

In addition, the proposed method can also be modified to suit various practical needs. For example, we may use a distributed learner \citep{Duan2019}  to overcome the communication barrier in the first step. This communication barrier comes from the fact that the datasets from different owners (e.g., hospitals) cannot be pooled on a single machine due to privacy regulations (e.g., HIPAA on sharing EHRs). Another possible extension is to learn an individualized treatment rule (ITR). In this setting, the first step can be implemented on observational data serving as ``real-world evidence,'' and the calibration can be done using clinical trial data. As such, a learned ITR can take advantage of volumes of observational data and retain robustness against possible confounding in observational studies.
 
\begin{center}
	{\large\bf Supplemental Materials}
\end{center}

\begin{description}
	
	\item[] \hspace{.65cm} Proofs of all theorems and additional simulation results are contained in the online supplemental materials. 
		
\end{description}

\singlespacing
\bibliographystyle{apalike}
\bibliography{ref.bib}
\end{document}